\documentclass[review]{elsarticle}

\makeatletter
\def\ps@pprintTitle{%
  \let\@oddhead\@empty
  \let\@evenhead\@empty
  \let\@oddfoot\@empty
  \let\@evenfoot\@oddfoot
}
\makeatother

\usepackage[english]{babel}
\usepackage{tabularx}
\usepackage{seqsplit}
\usepackage[letterpaper,top=2cm,bottom=2cm,left=3cm,right=3cm,marginparwidth=1.75cm]{geometry}
\usepackage[font=small,labelfont=bf]{caption}
\usepackage{amsmath}
 \usepackage{subfigure}
\usepackage{graphicx}
\usepackage[colorlinks=true, allcolors=blue]{hyperref}
\usepackage{tikz}
\usepackage{amssymb}
\usepackage[math]{anttor}
\usepackage{caption}
\usepackage{xcolor}
\usepackage{multirow}
\usetikzlibrary{shapes.geometric, arrows}
\tikzstyle{startstop} = [rectangle, rounded corners, minimum width=3cm, minimum height=1cm,text centered, draw=black, fill=red!30]
\tikzstyle{io} = [trapezium, trapezium left angle=70, trapezium right angle=110, minimum width=3cm, minimum height=1cm, text centered, draw=black, fill=blue!30]
\tikzstyle{process} = [rectangle, minimum width=3cm, minimum height=1cm, text centered, draw=black, fill=orange!30]
\tikzstyle{decision} = [diamond, minimum width=3cm, minimum height=1cm, text centered, draw=black, fill=green!30]
\tikzstyle{arrow} = [thick,->,>=stealth]

\begin{document}
\begin{frontmatter}

\title{Designing architectured ceramics for transient thermal applications using finite element and deep learning}
\author[1,2]{Elham Kiyani}
\author[3]{Hamidreza Yazdani Sarvestani}
\author[3]{Hossein Ravanbakhsh}
\author[2,4]{Razyeh Behbahani}
\author[5]{Behnam Ashrafi}
\author[3]{Meysam Rahmat}
\author[2,4,6]{Mikko Karttunen}
\cortext[cor1]{Corresponding authors. Emails: mkarttu@uwo.ca and hamidreza.yazdani@nrc.ca}
\address[1]{Department of Mathematics, The University of Western Ontario, 1151 Richmond Street, London, ON N6A~5B7 Canada}
\address[2]{The Centre for Advanced Materials and Biomaterials (CAMBR), The University of Western Ontario, 1151 Richmond Street, London, ON N6A~5B7 Canada} 
\address[3]{Aerospace Manufacturing Technology Centre, National Research Council Canada, 5145 Decelles Avenue, Montreal, QC H3T~2B2 Canada} 
\address[4]{Department of Physics and Astronomy, The University of Western Ontario, 1151 Richmond Street, London, ON N6A~3K7 Canada}
\address[5]{Aerospace Manufacturing Technology Centre, National Research Council Canada, 1200 Montreal Road, Ottawa, ON K1A~0R6 Canada}
\address[6]{Department of Chemistry, The University of Western Ontario, 1151 Richmond Street, London, ON N6A~5B7 Canada}

\begin{abstract}
Topologically interlocking architectures can generate tough ceramics with attractive thermo-mechanical properties. This concept can make the material design pathway a challenging task, since modeling the whole design space is neither effective nor feasible. We propose an approach to design high-performance architectured ceramics using machine learning (ML) with data from finite element analysis (FEA). 
Convolutional neural networks (CNNs) and Multilayer Perceptrons (MLPs) are used as the deep learning approaches.
A limited set of FEA simulation data containing a variety of architectural design parameters is used to train our neural networks, including learning how independent and dependent design parameters are related. A trained network is then used to predict the optimum structure from the configurations. A FEA simulation is run on the best predictions of both MLP and CNN algorithms to evaluate the performance of our networks. Although a limited amount of simulation data are available, our networks are effective in predicting the transient thermo-mechanical responses of possible panel designs. For example, the optimal design after using CNN prediction resulted in $\approx \! 30\%$ improvement in terms of edge temperature.

\end{abstract}
\end{frontmatter}

\section*{Highlights}

-High-performance architectured ceramics using ML with data from FEA are developed.

-MLP and CNN algorithms are used to evaluate the performance of the ML-FE model.

-The developed ML-FE model is effective in predicting the transient thermo-mechanical responses.

\section{Introduction}

The development of ceramic materials with better
thermo-physical properties is critical 
for a wide range of applications~\cite{ashby2013designing, ritchie2011conflicts}. A common practical strategy to address this challenge is to manipulate the ceramic architecture \cite{barthelat2015architectured, brechet2013architectured, sarvestani2022architectured}, an approach that nature has perfected~\cite{meyers2008biological}. In particular, topologically interlocking  stiff building blocks of well-controlled geometries provides a natural mechanism for achieving stability, specific thermo-mechanical properties, non-linear deformations and delayed localization in ceramics~\cite{fatehi2021accelerated, dyskin2003topological}. The design space for topologically interlocking ceramics is, however, vast, and finding the optimal architectures for given loading configurations remains a challenge. Similarly, predicting their 
performance 
is difficult as there are a multitude of design parameters, 
and even the mechanics of interlocking is complex. Therefore, a novel strategy is required to tune the performance of topologically interlocking ceramics and to predict and design their properties.

Deep learning and other ML techniques have undergone rapid development over the past few years, and they are being successfully applied to accelerate research in a multitude of areas including materials and structure design in engineering~\cite{so2019simultaneous, dimiduk2018perspectives, singh2020design, tan2020deep, behbahani2022machine, ravanbakhshcombining, patel2020predicting}.
One of the strengths of deep learning is its
capability of expressing 
strong nonlinear relationships~\cite{hornik1989multilayer,cybenko1989approximation}. 
In addition, the performance of deep learning methods is generally superior to other 
well-known 
ML techniques~\cite{zhang2021language, sun2020automatically, deldjoo2016content, brunton2020machine, rozenwald2020machine}.

Multilayer perceptron (MLP) and convolutional neural networks (CNNs) are deep learning methods that have recently been proven to be very effective 
in image processing and traffic sign recognition~\cite{matsumoto1990several, shustanov2017cnn, huang1999mlp}.
MLPs are neuronal networks (NNs) with one or more hidden layers. They have been found to be particularly helpful for extracting higher-order statistics. 
In addition to being used for identifying discrete classes, MLPs are also widely used for predicting continuous values in regression~\cite{pattanayak2021application, kiyani2022machine, keybondorian2017application, murtagh1991multilayer}.
These models are typically feedforward artificial neural networks consisting of a series of layers. In each layer, the weighted sum of the inputs is calculated and then an activation function is applied to assign a single value to the next neuron~\cite{barlow1995feed}.

CNNs are mainly used for image recognition; since the input is viewed as an image, an architecture can be built to extract the properties of the image. Such may include edge detection, gradient recognition, and smoothing, for example. Convolution is a filter that is able to perform complex operations on the images. CNNs offer several advantages, notably, great precision and accuracy particularly when dealing with problems that evolve over time~\cite{sun2020automatically, sultana2018advancements}.

In this work, deep learning using CNN and MLP frameworks
was employed to achieve optimal designs of architectured interlocked ceramic panels.
Initial training was done
using finite element analysis (FEA) data. The three different tile designs based on truncated tetrahedral designs were then investigated for their thermal performance under thermal shock loading. The aim was to
find optimal designs
in terms of energy dissipation, energy absorption, friction between tiles, and resistance to failure. 
The predicted optimized designs were then validated via FEM.

The two main outcomes of this work are 1) MLP- and CNN-based deep learning models that are capable of learning relations between the parameters in architectured ceramics under thermo-mechanical loads, and thus assist in designing optimal architectures for 
thermo-mechanical applications of varying demands. 2) 
We use
the transient time-dependent behaviour of the architectured systems under thermo-mechanical loads, that is,
the performance of these complex systems is monitored over a period of time during and after the thermal shock. Time is fed into the networks as an independent parameter
making the rest of the parameters functions of time. 
To the best of the authors' knowledge,
such an approach with
transient thermo-mechanical problems with several time-dependent parameters has not been previously reported in the literature.
Furthermore, this 
also appears to be
the first attempt to use CNNs to design architectured ceramics.. 
In this study, neural networks are trained implemented using the TensorFlow framework~\cite{tensorflow2015-whitepaper} and Comsol Multiphysics® was used for generating FEM simulation data. 

The rest of the paper is organized as follows. Data generation and structural design are discussed in Section~\ref{Data generation, structural design}. An overview of deep learning and the proposed models based on MLPs and CNNs, are presented in Section~\ref{Deeplearning}, and finally, Section~\ref{Conclusion} presents conclusions.

\section{Data generation and structural design} \label{Data generation, structural design}

\subsection{Structural Design}

The construction of topologically interlocking ceramics from truncated tetrahedral blocks is shown in Fig.~\ref{fig:Fig1}. The basic building block is obtained by truncating a non-regular tetrahedron at the median plane and at a plane at a distance $H$ from the median plane. The resulting block has six faces, including two parallel surfaces at the top and the bottom, two facing surfaces tilted inward and the other two surfaces tilted outward by the same or different interlocking angle $\theta$, Fig.~\ref{fig:Fig1}. 

The lower face of the building block is an $l \times l$ square, while the top face is a rectangle of dimensions $l+ 2H\tan \theta_{2}$ by 
$l-2H\tan\theta_{1}$. The architecture of the building block is, therefore, fully represented by the three independent parameters $H$, $l$, and $\theta$. Based on these independent geometrical parameters, three topologically interlocking ceramic systems were designed: (I) constant values for the interlocking angle and the tile size, (II) constant interlocking angle and variable tile size, and (III) variable values for both the interlocking angle and the tile size. The tile thickness, $H$, was kept constant in all the three cases.

\begin{table}
\begin{tabular}{|l|l|l|}
\hline
Category & Simulation Parameter & Value \\
\hline
\hline
\multirow{2}{*}{Geometry} & Panel size  & $50$ mm \\ \cline{2-3}
 & Panel thickness (H)  & $2.54$ mm \\ \cline{2-3}
\hline
\multirow{10}{*}{Material properties} & Tensile strength ($\sigma_{ts}$) & $220$ MPa  \\ \cline{2-3}
 &  Compressive strength ($\sigma_{cs}$)  & $2068$ MPa \\ \cline{2-3}
  & Density ($\rho$)  & $3800$ kgm$^{-3}$ \\ \cline{2-3}
   & Thermal conductivity (K)  & $24.6$ W m$^{-1}$ K$^{-1}$\\ \cline{2-3}
    & Heat capacity ($C_p$)  & $880$ Jkg $^{-1}$ K$^{-1}$ \\ \cline{2-3}
     & Young’s modulus (E)  & $303$  GPa  \\ \cline{2-3}
      & Poisson’s ratio ($\nu$)  & 0.21 \\ \cline{2-3}
       & Coefficient of thermal expansion ($\alpha$)  & $8.28 \times 10^{-6}$ K$^{-1}$  \\ \cline{2-3}
        & Surface emissivity ($\epsilon$) & $0.4$ \\ \cline{2-3}
 & Heat transfer coefficient (h) & $10$ W m$^{-2}$ K$^{-1}$ \\ 
\hline
\multirow{5}{*}{Contact parameters} & Gap conductance ($h_{g}$)  & $2000$ W m$^{-2}$ K$^{-1} $ \\ \cline{2-3}
 & Surface roughness, asperities average height ($\sigma_{asp}$) \cite{beausoleil2020deep, esmail2021engineered}  & $10 \, \mu \mathrm{m}$ \\ \cline{2-3}
  & Surface roughness, asperities average slope ($m_{asp}$) \cite{beausoleil2020deep, esmail2021engineered}  & $0.0167$ \\ \cline{2-3}
   & Vickers correlation coefficient ($c_1$)  & $10.5$  GPa \\ \cline{2-3}
 & Vickers size index ($c_2$)  & $-0.03$ \\ 
\hline
\multirow{5}{*}{Mesh properties} & Type of predefined mesh & Finer \\ \cline{2-3}
& Maximum element size & $1.43$ mm   \\ \cline{2-3}
 & Minimum element size  & $0.104$ mm  \\ \cline{2-3}
  & Maximum element growth rate  & $1.4$\\ \cline{2-3}
   & Curvature factor & $0.4$ \\ \cline{2-3}
 & Resolution of narrow regions & $0.7$\\ 
\hline
\end{tabular} 
\caption{\label{tab:parameters}Simulation parameters.}
\end{table}

\subsection{Finite element modelling}

Comsol Multiphysics® was employed for the finite element analysis of interlocking ceramics under thermal shock. Thermal loading was applied at the centre of the panel in the form of temperature increase over a circular surface with a radius ($R$) of $7.5$ mm. Depending on the configuration of the tiles and their sizes, one or more tiles could be exposed to direct temperature increase. The loading profile was defined 
as follows to represent a thermal shock: 
A temperature ramp of $97.5^{\circ}\mathrm{C}/s$ starting at $25^{\circ} \mathrm{C}$ (at $t = 5$ s) and ending at $1000^{\circ} \mathrm{C}$ (at $t = 15$\,s) was first applied. This was followed by a hold at 
$1000^{\circ}\mathrm{C}$ until $t = 35$ s. Then, the thermal load was removed and the system was let to cool down naturally via thermal convection and radiation. 

\begin{figure}
    \centering
    \begin{tabular}{c}
        \includegraphics[width=0.9\textwidth]{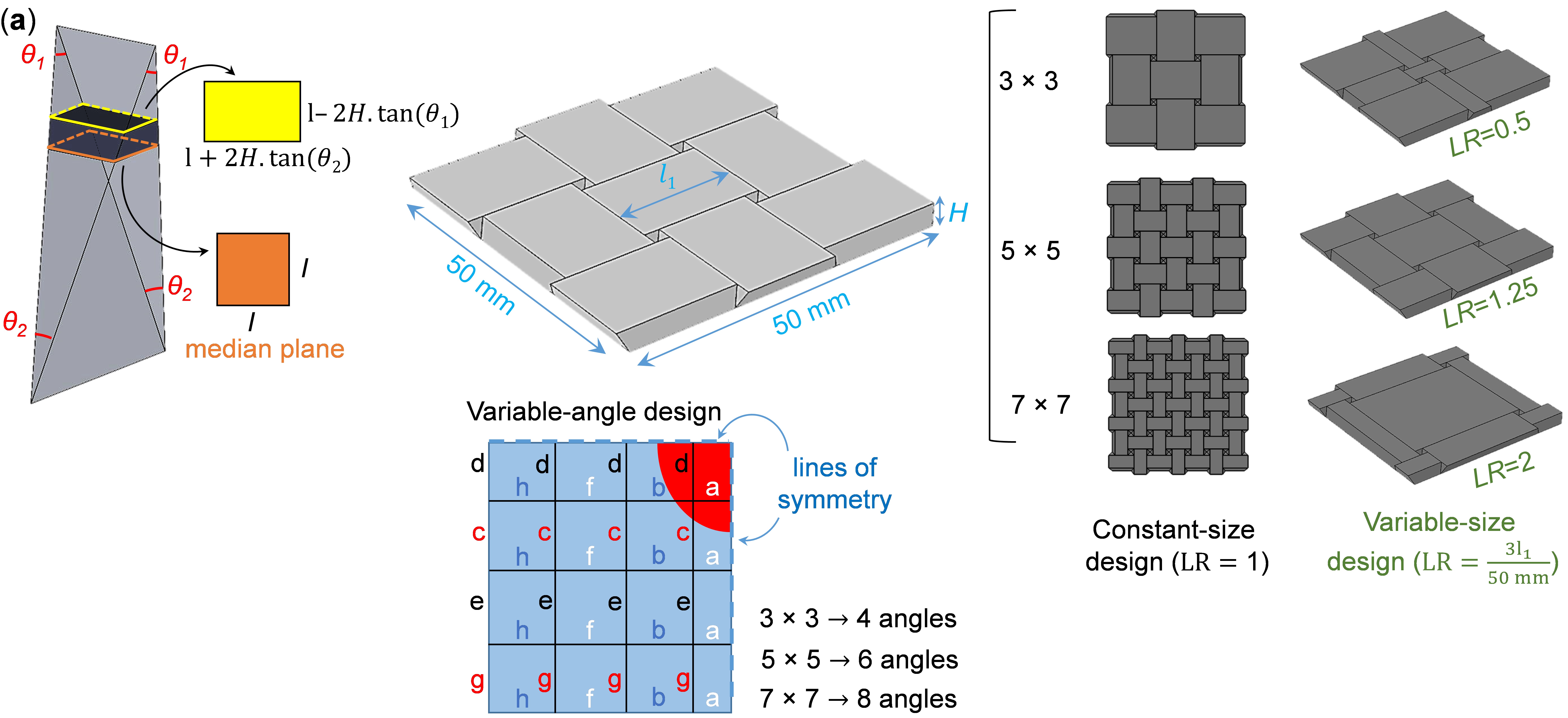} \\
        \includegraphics[width=0.55\textwidth]{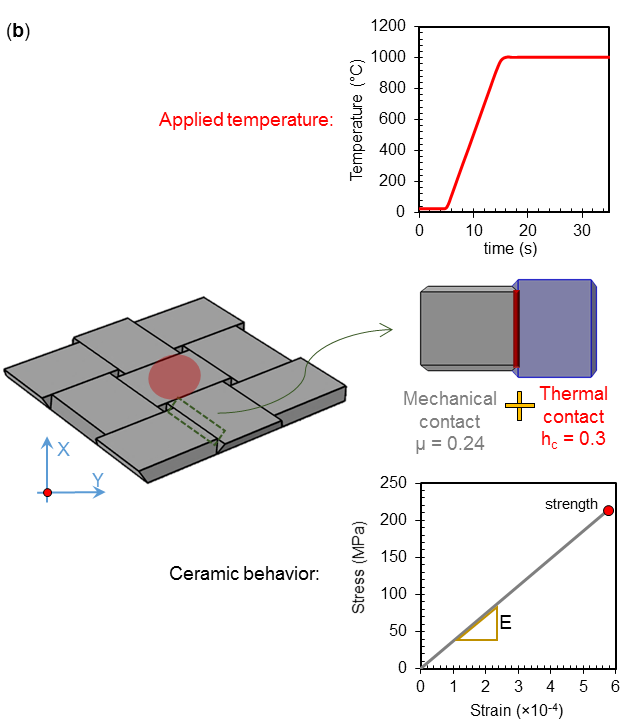} \\
    \end{tabular}
    \caption{Design of topologically interlocking panels: (a) tiles with constant interlocking angle and size, constant interlocking angle and variable size, and variable interlocking angle and size. Each letter (a, b, c, etc.) represents different interlocking angles of each tile for the case of variable-angle design. (b) FE analysis schematic: Ceramic behaviour, material properties, and applied temperature. The red line represents thermo-mechanical contact properties between the tiles. $\mu$  and $h_c$ are the friction and heat transfer coefficients, respectively.}
    \label{fig:Fig1}
\end{figure}

Both mechanical and thermal contacts were defined between the tiles. A Coulomb friction model with a constant friction coefficient of $0.24$ was implemented between the paired surfaces as the mechanical contact. The ceramic was modeled as an isotropic linear elastic material. Constriction conductance model with interstitial gas was employed to simulate the thermal contact between the tiles. For the constriction conductance, $h_c$, the Cooper-Mikic-Yovanovich correlation \cite{bejan2003heat} was used. The initial temperature of the structure as well as the environment temperature were set to $25^{\circ}\mathrm{C}$. All the parameters related to contact and material properties are listed in Table~\ref{fig:Fig1}. 

The top and the bottom surfaces of the panel were modelled as free surfaces, and a fixed boundary condition was applied to the edges. Due to the symmetry of the architectured structure, a quarter of the panel was simulated to reduce the computational cost. Since the contact problem is highly nonlinear and may cause issues with numerical convergence, an initial spring foundation (with a spring constant of $K_v$) was defined for the entire ceramic panel. This value, which was only selected to assist with simulation convergence, was reduced to zero over time before thermal loading started at $t = 5$ s. It should be mentioned that some of the parameters are Comsol Multiphysics® specific. This includes a predefined mesh, maximum element growth rate, curvature factor, and the resolution of narrow regions.

\section{Deep learning}\label{Deeplearning}

Deep learning tools were employed to predict the interlocking designs 
for optimal thermo-mechanical output. We present two frameworks for learning the thermo-mechanical response to a thermal shock: (I) An MLP network, and (II) CNN.  In an MLP, each layer consists of a set of neurons, which are connected to those in the previous layer. CNNs are, however, structured differently in contrast to a regular neural network. They contain three-dimensional layers
which are characterized by width, height, and depth. Only a small portion of neurons in a given layer are connected to neurons in the previous layer~\cite{hastie2009elements}.

\begin{figure}
\centering
\includegraphics[width=\textwidth]{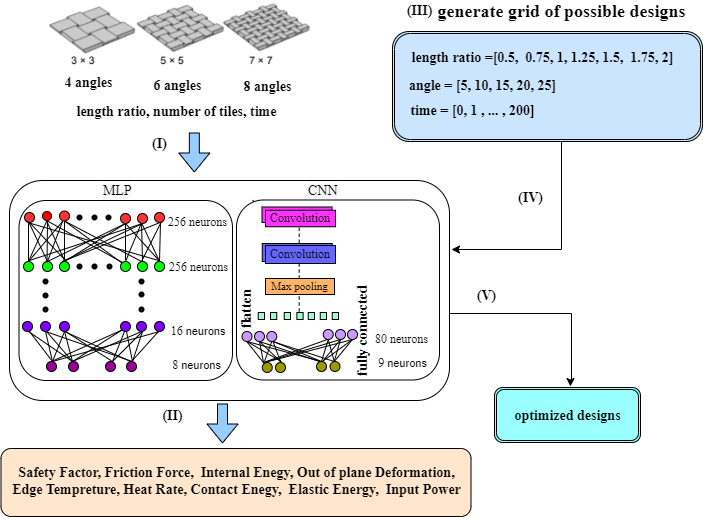}
\caption{
The general steps in discovery of optimized designs. (I) An MLP and CNN with seven inputs for the ($3\times3$) case, nine and eleven inputs for the ($5\times5$) and ($7\times7$) cases consisting of angles (4 angles for ($3\times 3$), $6$ angles for ($5\times 5$) and 8 angles for ($7\times7$)), length ratio,  number of tiles ($9$, $25$,  and $49$ for the three cases, respectively) and simulation time steps that are $[0, 600]$. The input parameters are fed through the networks at the first step. (II) MLP and CNN networks are trained to predict outputs consisting of the safety factor, friction force, internal energy, out-of-plan deformation, edge temperature, heat rate, contact energy, elastic energy, and input power. (III) Generation of all potential designs. Possible values for length ratio are
($[0.5, 0.75, 1, 1.25, 1.5, 1.75, 2]$), for the angles they are ([$5^{\circ}, 10^{\circ}, 15^{\circ}, 20^{\circ}, 25^{\circ}$]), and for time ($[0 s,.., 200 s]$). (IV) The generated possible designs are fed through the trained MLP and CNN networks to predict the thermo-mechanical response to a thermal shock. (V) The optimal set of interlocking parameters is determined by predicting parameters based on the grid possible designs.}
\label{fig:Schematic}
\end{figure}

Fig.~\ref{fig:Schematic} shows the steps for finding optimized designs for ceramic panels. Tile length ratios, contact angles, and the number of tiles are the geometric variables that are fed through the MLP and CNN networks.
For the ($3\times 3$) case, the number of tiles is $9$ and the input consists of four angles. For the ($5\times 5$) and ($7\times7$) cases, the number of tiles are $25$ and $49$, and the networks are trained with $9$ and $11$ inputs comprising $6$ and $8$ angels, respectively. A tile's contact angle can be set at $5, 10, 15, 20, 25$ degrees and the length ratio (LR) can be $0.5, 0.75, 1, 1.25, 1.5, 1.75, 2$. Length ratio for a $3 \times 3$ structure is defined as the ratio of the size of the middle tile to one-third of the total size of the ceramic panel. Accordingly, $\mathrm{LR} = 1$ represents the case in which all the tiles are of the same size, and $\mathrm{LR} > 1$ represents the situation in which  the middle tile is larger than the peripheral tiles. 

The outputs of the networks include the panels' safety factor, friction force, internal energy, out-of-plane deformation, edge temperature, heat rate, contact energy, elastic strain energy, and input power over the simulation period:
\begin{itemize}
\item Friction force: When a building block expands, a lateral force is applied to the adjacent tiles. Owing to the unique interlocking architecture of the tiles, this lateral force causes the tiles to slide over each other. 
\item Input power: Thermal energy is applied to the architectured ceramic over a circular region at the bottom surface of the structure. 
\item out-of-plane deformation: Upon applying thermal loading, ceramic tiles experience thermal expansion, resulting in sliding over the adjacent tiles. Due to the architecture of the ceramics, the tiles’ thermal expansion leads to out-of-plane deformation. 
\item Safety factor: The minimum safety factor throughout the structure was calculated using the Drucker-Prager damage index in FEA, which is a built-in function in Comsol Multiphysics®. 
\item Internal, contact, and elastic energies: The net thermal energy that contributes to enhancing the internal energy. The thermodynamic work done by the system consisting of the elastic strain energy in tiles, the contact energy between the tiles, the work needed to overcome friction, and the work done on the environment. 
\item Edge temperature: The capability of tuning thermal conduction is an important design parameter for architectured ceramic panels. The temperature of the edges is an indicator of how the panels transfer heat when temperature is increased at the center of the structure. 
\item Heat rate: The net heat rate is calculated based on the difference between the input thermal energy and the output.
\end{itemize}

The details of MLP and CNN architectures are discussed in Section~\ref{subsec:MLPandCNN}.  
The design space of interlocking architectured ceramics includes tiles with different angles and sizes. The performance and characteristics of a given design are determined via output parameters that are discussed in the following sections.

\begin{figure}
\centering
\includegraphics[width=\textwidth]{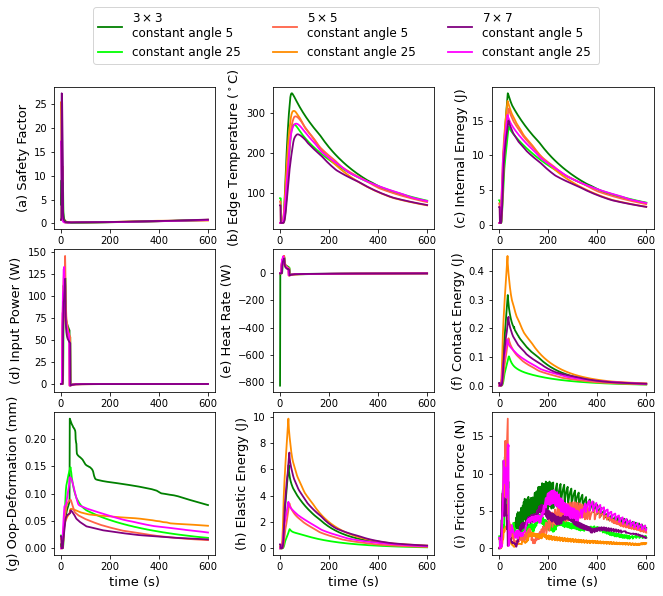}
\caption{Time-domain response of constant-angle interlocking ceramic designs with $3 \times 3$, $5 \times 5$, and $7 \times 7$ array of tiles: (a) Safety Factor, (b) edge temperature, (c) internal energy, (d) input power, (e) heat rate, (f) contact energy, (g) out-of-plane deformation, (h) elastic energy, and (i) friction force.
}
\label{fig:time_plot}
\end{figure} 


\subsection{Multilayer Perceptron (MLP) and Convolutional neural network (CNN) performance}\label{subsec:MLPandCNN}

MLPs can be used to learn mappings between inputs and outputs. They are comprised of multiple hidden layers and single input-output layers. A layer consists of neurons that are connected by weight-bias parameters. The architecture of an MLP is crucial to the success of the network. The best architecture for a given problem can never be specified with exact precision in advance, and different problems require different architectures.
Generally, feedforward involves moving forward with the inputs and weights (assumed in the first run) until the output is reached. The back-propagation algorithm is the most widely used algorithm for designing MLPs. As the name implies, backward propagation moves from output to input. In order to minimize a loss function, backward propagation is used to adjust and correct the weights~\cite{zerguine2001multilayer, yu2022s2}.

The MLP architecture for learning the thermal shock parameters consists of an input layer with seven, nine, and eleven neurons (contact angles, length ratio, number of pieces, and time) for the cases of ($3 \times 3$), ($5 \times 5$), and ($7 \times 7$), respectively. These neurons are connected to the hidden layers with $256/ 256/ 256/ 128/ 64/ 32/ 16/ 8$ neurons each. In the output layer, a dense layer with  $9$ neurons corresponding to the thermo-mechanical response to a thermal shock is used. The network is trained for $20,000$ epochs using the ADAM optimizer \cite{diederik2014adam}, rectified linear unit (ReLU) activation function \cite{brownlee2019gentle}, and the mean squared error (MSE) as the loss function. 
The augmented data set is divided into two $80\%$ and $20\%$ sets for training and test.   

On the other hand, a CNN consists of layers of neurons that learn hierarchical representations. There is an input layer followed by an output layer, and in-between, there are hidden layers that transform the input's feature space into the output. Several optimizable filters are used in convolutional layers to transform the input or previous hidden layers~\cite{kattenborn2021review}. The mathematical details, operations and functionality of the layers are beyond the scope of this paper and can be found in Ref.~\cite{rawat2017deep}.

The CNN structure proposed in this work has two convolutional layers (Conv1D) with $256$ and $32$ filters as well as the kernel size $3$ followed by a 1D max pooling operation with the pool size $2$. A flat layer is followed by one dense layer with $80$ neurons and ReLU~\cite{brownlee2019gentle} as an activation function, and a dense layer with $9$ neurons corresponding to the outputs. 
The CNN model was trained for $20,000$ epochs using the ADAM optimizer and the ReLU activation function, and mean squared error (MSE) as the loss function. 

\begin{figure}
    \centering
    \includegraphics[width= 1\textwidth]{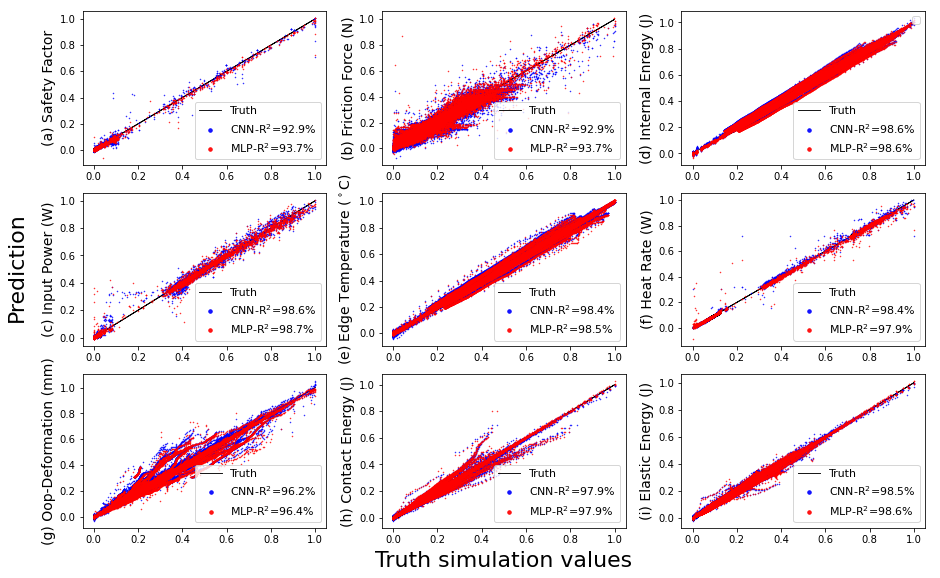}
    \caption{Performance of the MLP and CNN parameter  prediction. Each plot shows the true FEA simulation values vertically and the predictions generated by MLP and CNN horizontally. The red and blue dots represent the MLP and CNN predictions, respectively. The predicted points are all close to the diagonal line, confirming the power of MLP and CNN networks in predicting parameters. $R^2$ (see Equation~\ref{eq:R2}) for each parameter is higher than $92.9 \%$.
    }
    \label{fig:method_comp}
\end{figure}

A comparison of the frameworks' performances for a random train and test split is summarized in Fig.~\ref{fig:method_comp}.
The figure shows the test set of output parameters and the corresponding predictions by MLP and CNN. Results are mainly on the diagonal line, which indicates a good performance.
The coefficient of determination, $R^2$ which was used to compare the performances of the networks, is reported for each prediction. It is worth noting that $R^2$ is expressed as
\begin{equation}\label{eq:R2}
R^2 = 1- \frac{\sum_{i=1}^{n}(y_{i} - \widehat{y_i})^2}{\sum_{i=1}^{n}(y_{i} - \overline{y})^2},
\end{equation}
where $y_i$ is the observed data, $n$ is the number of data, $\overline{y}$ is the mean value of the observed values, and $\widehat{y_i}$ is the predicted data.  
As marked in Fig.~\ref{fig:method_comp}, all $R^2$ scores are higher than $92 \%$, which shows that the model is trained well.

\begin{figure}
    \centering
\includegraphics[width= 1\textwidth]{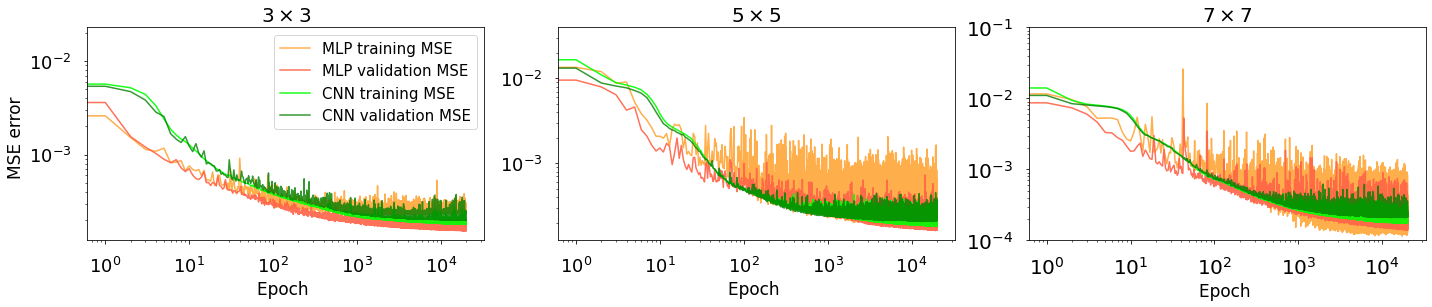}
    \caption{Trace of MSE (see Equation~\eqref{eq:MSE}) errors for MLP and CNN networks. The light and dark orange lines represent the error on the MLP training and validation sets as a function of epochs. The light and dark green lines correspond to the error on the CNN training and validation sets. Learning curves show that the training and validation curves are very similar as all decrease to a point of stability.}
    \label{fig:MSE}
\end{figure}

Another factor that can be used to evaluate networks' performances is MSE. It is the most commonly used loss function, and can be expressed as the mean of the squared differences between true and predicted values as
\begin{equation}\label{eq:MSE}
     \mathrm{MSE} = \frac{1}{n} \sum_{i=1}^{n}(y_{i} - \widehat{y_i})^2,
\end{equation}
where $y_i$ and $\widehat{y_i}$ are the observed and predicted values, and $n$ is the number of data points. To evaluate the networks' performances, the MSE as a function of epochs is given 
in Fig.~\ref{fig:MSE}.
As the figure shows, the MSE values are small ($\sim \! 10^{-4}$), indicating that the target parameters (safety factor, edge temperature, internal energy,  input power, heat rate, contact energy, out of plan deformation, elastic energy, and friction force) learned by the proposed MLP and CNN networks are close to the true FEA-simulated ones. The figure also shows that the MSE errors for both training and validation loss decrease to a point of stability with a minimal gap between the two final loss values.

After confirming the performance of the models, two 
case studies were considered. In the first one, a panel design giving the lowest maximum edge temperature and maximum out-of-plane deformation under a thermal shock was required. This scenario is representative of thermal shield ceramics for which  minimum temperature at the edges and the highest geometric stability under thermal load is required. In the second case study, the aim is to design a structure with the highest maximum internal energy but the lowest maximum elastic strain energy. This scenario is a representative of heat sink applications, where maximum amount of absorbed heat, while maintaining minimum thermal stress is desirable.
The rest of this paper is organized according to these two scenarios, and the details of the MLP and CNN models are discussed only for four output parameters: internal energy, out-of-plane deformation, edge temperature, and elastic strain energy.

\begin{figure}
\centering
\includegraphics[width=\textwidth]{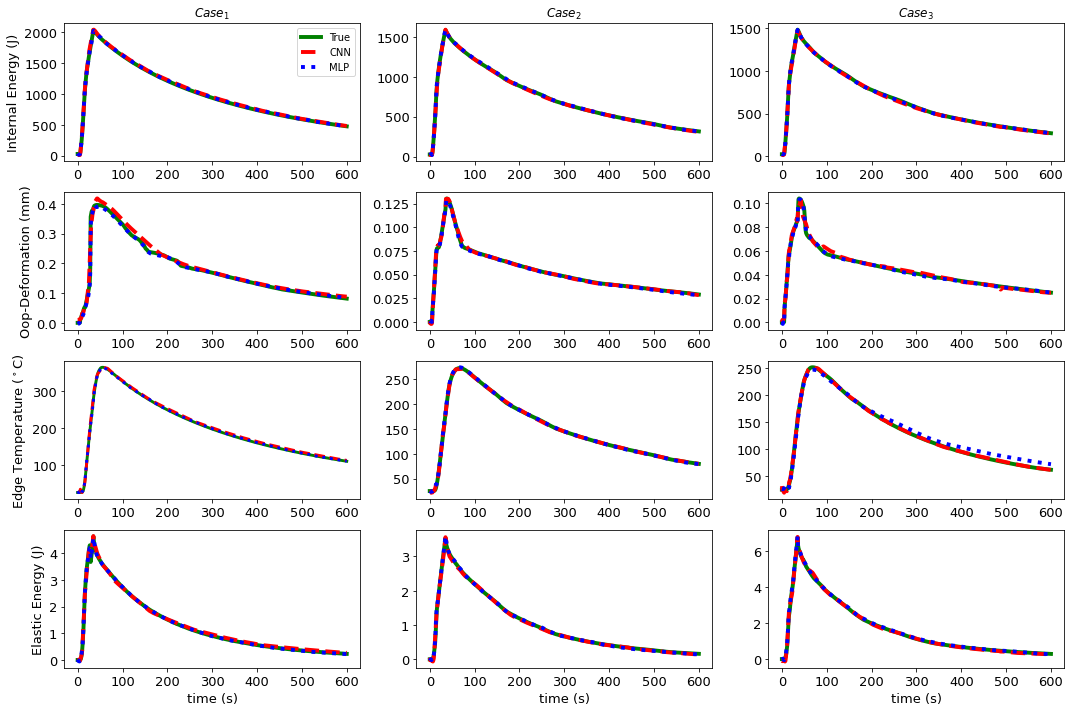}
\includegraphics[width=\textwidth]{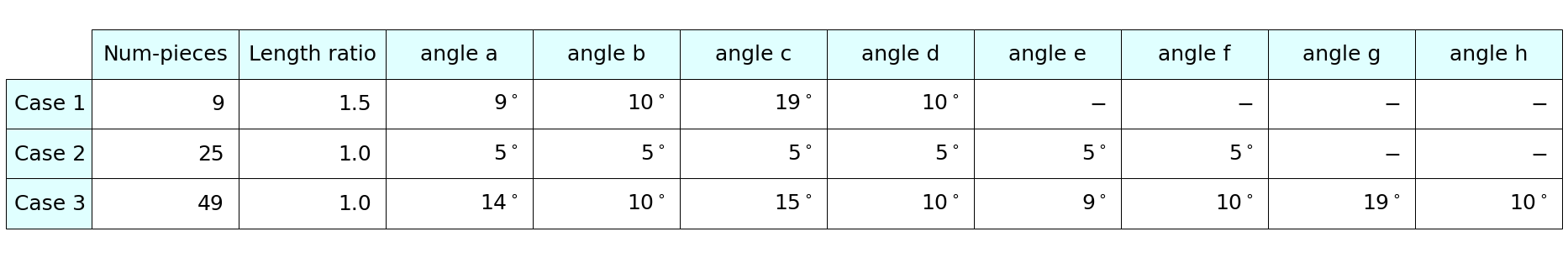}
\caption{Comparison of MLP and CNN performance on prediction of four sets of outputs:  the panel's minimum out-of-plane deformation, maximum internal energy, minimum elastic strain energy, and the tiles' highest edge temperature over the simulation period. In each plot, the horizontal axis shows the time and the vertical axis the outputs. To demonstrate the ability of MLP and CNN in predicting parameters, three randomly selected cases are presented (three columns of charts) in the table.}
\label{fig:MLP and CNN performance}
\end{figure}

The network performances and comparison with the true values (FEA simulation results) are
shown in Fig.~\ref{fig:MLP and CNN performance}. As examples, three different cases were randomly selected and are reported in the table below the figures. The figures on the left column represent predictions for a ($3 \times 3$) case, and the centre and the right columns show the predictions for  the cases of ($5 \times 5$) and ($7 \times 7$), respectively. 
The data demonstrates that the MLP and CNN models have excellent accuracy and precision.

\begin{figure}
\centering
\includegraphics[width=\textwidth]{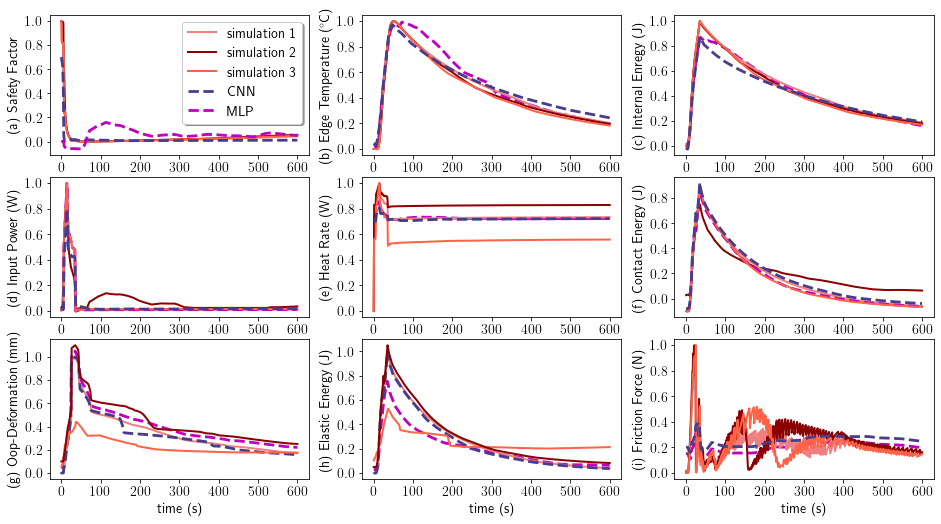}
\caption{Comparison of the CNN and MLP predictions and multiple FEA simulations using a constant angle of $10^\circ$ for a ($3 \times 3$) case using different meshing (intentionally done). The pink, and the light and dark red lines
show data from the independent simulations. 
The CNN and MLP predictions are shown in dark blue and purple dashed lines, respectively.  
When there is a gap between simulation re-runs, both MLP and CNN predict values that are between simulation results.}
\label{fig:repeated-3by3}
\end{figure}

To evaluate the consistency and reproducibility of the FEA simulation results, some of the simulations were repeated and all the result sets were used for training the networks. Constant angle of $10^\circ$ for a ($3 \times 3$) case was selected as an example to present in  
Fig.~\ref{fig:repeated-3by3}. The figure illustrates that when gaps exist between the re-runs, the MLPs and CNNs predictions fall in-between the FEA simulation results. After training the networks (see Fig.~\ref{fig:Schematic} for the process), all ceramic panel designs were then fed to the trained networks to predict the corresponding output parameters. 

\subsection{Generating a grid of possible ceramic panel  designs}

To design high-performance architectured ceramics, a grid composed of all possible designs was generated. The input parameters  included length ratios of
($[0, 0.75, 1, 1.25, 1.5, 1.75, 2]$), angles ([$5^{\circ}, 10^{\circ}, 15^{\circ}, 20^{\circ}, 25^{\circ}$]), and time range of ([$0\,\mathrm{s},600\,\mathrm{s}$]).  
By considering all possible time values within the $600$\,s range in $1$\,s increments, the number of possible designs in the grid for the ($ 3 \times3$) case is $((2625000, 7)$, for ($ 5 \times5$) it is  $(65625000 , 9 )$ and for the ($ 7 \times7$) it is $(1640625000 , 11 )$. 
Computational restrictions, including memory limitations, made it impossible to predict all of the designs, especially for the  ($7 \times 7$) case. 
Based on Fig.~\ref{fig:MLP and CNN performance}, the optimized architecture designs
accrue before $t=200$\,s. Therefore, in this case, only grid values for $t=[0,200]$ were generated for the final optimizations. 

\begin{figure}
\centering
\includegraphics[width=\textwidth]{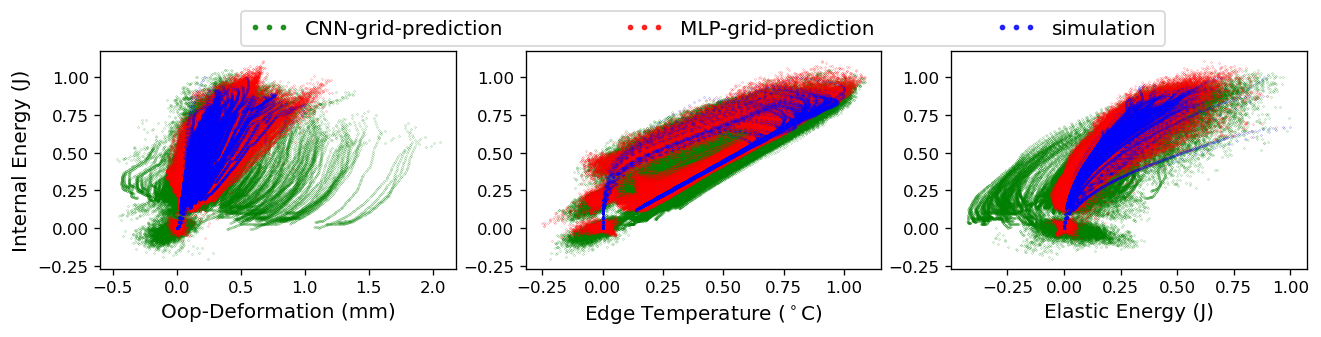}
\caption{Prediction of MLP and CNN on all possible designs in the generated grid for  the ($3 \times 3$) case. The CNN predictions are marked by green dots and the MLP predictions with red. The simulation results are presented with blue dots. The data show an excellent match between the CNN- and MLP-predicted parameters and the FEA simulation results.}
\label{fig:Grid-points}
\end{figure}

Using the workflow in Fig.~\ref{fig:Schematic}-IV, the generated designs were fed into the networks trained by the FEA simulation results. 
Optimised designs were derived based on the predicted outputs
of the possible designs. The MLP and CNN predictions on the generated grid of possible designs for ($3 \times 3$) are shown in Fig.~\ref{fig:Grid-points} as examples. 
The figure shows excellent agreement between the parameters predicted by CNN and MLP
with the FEA simulation values. 

\subsection{Architecture designs}

Two application scenarios were discussed earlier: thermal shields and thermal sinks. In this section, architecture designs for these two scenarios are presented. 
For each scenario, two important output parameters are highlighted. 

In the first scenario (i.e., thermal shielding), out-of-plane deformation and edge temperature are the crucial performance evaluation indicators. It should be noted that the importance of each of these parameter, is significantly influenced by the specific target application. For instance, when the edge temperature and out-of-plane deformation have equal importance, a $50-50 \%$ weight can be associated to each. The range of associated weights varies in pairs of $100 \%$ and $0 \%$ for each parameter.

\begin{figure}
    \centering
    {\includegraphics[width=0.44\textwidth]{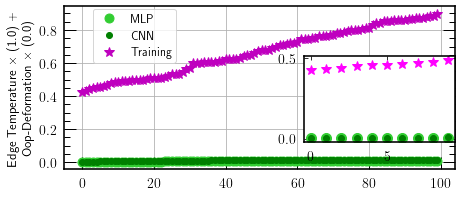}}
    {\includegraphics[width=0.44\textwidth]{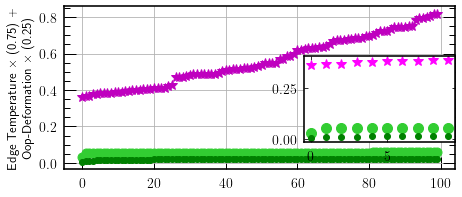}}
    {\includegraphics[width=0.44\textwidth]{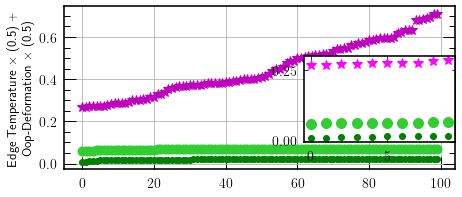}}
    {\includegraphics[width=0.44\textwidth]{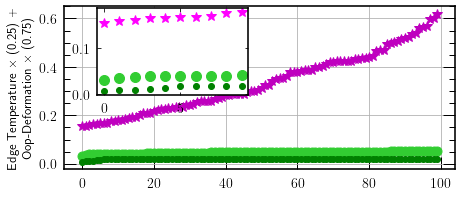}}
    {\includegraphics[width=0.44\textwidth]{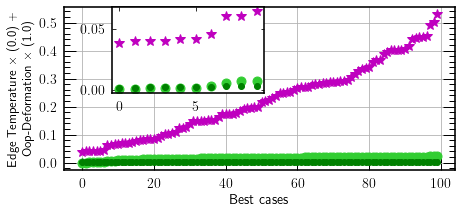}}
    \caption{First scenario (thermal shielding): The best performing cases sorted based on a weighted pair of the lowest maximum edge temperature and the lowest maximum out-of-plane deformation during the entire thermal shock cycle. The 100 best cases according to FEA simulation used in training (magenta stars), CNN prediction (dark green circle), and MLP predictions (light green circles) are shown. The first 10 cases (10 best cases) are shown in the inset.}
    \label{fig:ceramic-panel-designs1}
\end{figure}

The first scenario seeks panel designs with the lowest maximum edge temperature and out-of-plane deformation during the entire thermal cycle. The performances of all the possible designs in the grid were determined and the $100$ best ceramic panel designs
with a few combinations of weight pairs (different levels of importance for the two target output parameters) are shown in Fig.~\ref{fig:ceramic-panel-designs1}.  The first 10 cases are shown in the inset for more detail; the magenta stars indicate the $100$ most appropriate cases simulated by FEA, and the circles indicate the predictions from MLP and CNN.

The best performing cases (horizontal axes) in Fig.~\ref{fig:ceramic-panel-designs1} were sorted based on the weighted sum of the edge temperature $(100\%, 75\%, 50\%, 25\%, 0\%)$ and out-of-plane deformation with   
\seqsplit{$(0\%, 25\%, 50\%, 75\%, 100\%)$.}
The best design structure
turned out to be one of the  ($7 \times 7$) cases. 
The figure shows clearly 
that the performance of the best cases predicted 
by MLP and CNN are significantly superior to the limited FEA simulation results that were used for training.

\begin{figure}
    \centering
    {\includegraphics[width=0.44\textwidth]{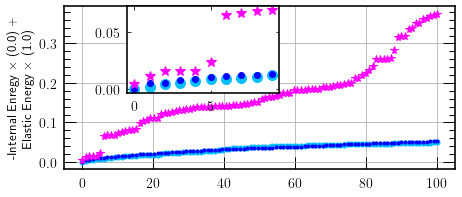}} 
    {\includegraphics[width=0.44\textwidth]{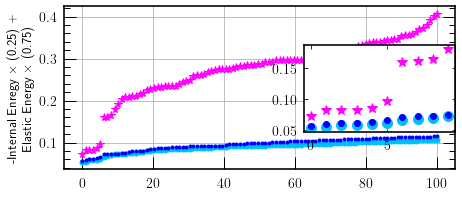}}
    {\includegraphics[width=0.44\textwidth]{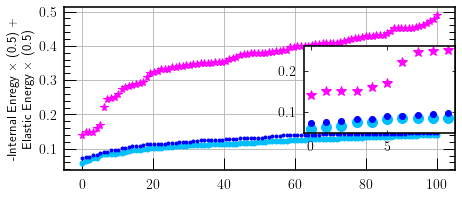}}
    {\includegraphics[width=0.44\textwidth]{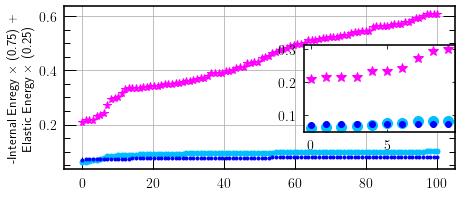}}
    {\includegraphics[width=0.44\textwidth]{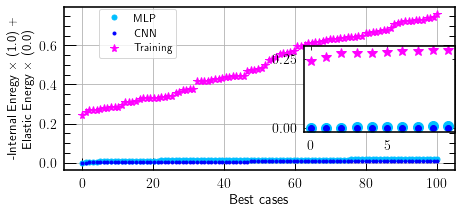}}
    \caption{Second scenario (thermal sink): The best performing cases sorted based on a weighted pair of the highest maximum internal energy and the lowest maximum elastic strain energy during the entire thermal shock cycle. The 100 best cases according to FEA simulation used in training (magenta stars), CNN prediction (green circle), and MLP predictions (blue circles) are shown. The first ten cases (10 best cases) are shown in the inset.}
    \label{fig:ceramic_panel_designs2}
\end{figure}

Similar approach for sorting the performances based on weighted pairs
was followed for the second scenario (thermal sink) with the difference that the output parameters
were the highest maximum internal energy and the lowest maximum elastic strain energy. The best performing designs are shown in Fig.~\ref{fig:ceramic_panel_designs2}. The magenta stars represent the $100$ FEA simulation results, and the cyan and green circles show the MLP and CNN predictions from the grid of all possible designs. It is worth noting that
the best predicted design structure is a ($3 \times 3$) case.

\begin{figure}
    \centering
    {\includegraphics[width=1\textwidth]{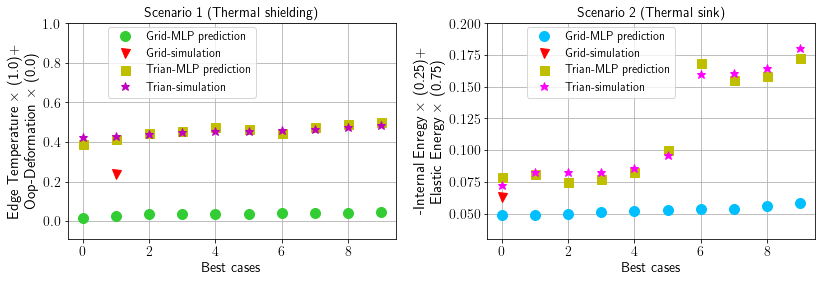}} 
    \caption{Results of FEA simulations and MLP predictions for both scenarios. First scenario is with the weighted pair of $1$ and $0$ for the edge temperatures and out-of-plane deformations, respectively. The second scenario is with the weighted pair of $75 \%$ and $25\%$ for internal energy and elastic strain energy. The magenta stars and olive green squares are FEA simulation results used during training and the corresponding predictions by MLP, respectively. 
    The green and blue circles show the MLP predictions of the 10 best performing cases in the grid (sorted as in Figs.~\ref{fig:ceramic-panel-designs1} and~\ref{fig:ceramic_panel_designs2}). The red triangles show the FEA simulation values based on the MLP predictions of the best cases (case 1 for the first scenario, and case 0 for the second scenario). The details of optimized designs for these cases are presented in Tables~\ref{table:first-scenario} and ~\ref{table:second-scenario}.}
    \label{fig:MLP-best-cases.png}
\end{figure}

\begin{figure}
\centering
\includegraphics[width=0.44\textwidth]{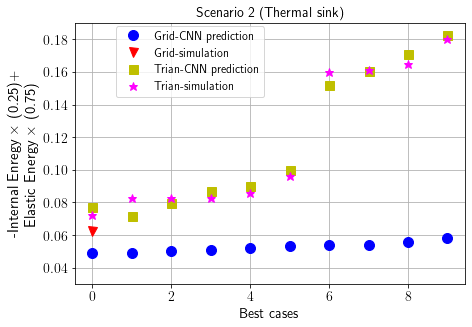}
\caption{Results of the FEA simulations and CNN predictions for the second scenario (thermal sink) with the weighted pair of $25 \%$ and $75\%$ for internal energy and elastic strain energy, respectively. The magenta stars and olive green squares are the FEA simulation results used during training and the corresponding predictions by CNN, respectively. The green circles show the CNN predictions of the 10 best performing cases in the grid (sorted similar to  Fig.~\ref{fig:ceramic_panel_designs2}). The red triangles shows the FEA simulation values based on the CNN prediction of the best case (case 0). The details of optimized designs
are presented in Table~\ref{table:second-scenario}.}
\label{fig:CNN-best-cases.png}
\end{figure} 

To evaluate the capabilities of the deep learning algorithms in capturing the behaviour of architectured ceramics under thermal shock, MLP predictions of the best performing cases for both of the above scenarios were
used in 
FEA simulations. 
Fig.~\ref{fig:MLP-best-cases.png} shows a combination of Figs.~\ref{fig:ceramic-panel-designs1} and~\ref{fig:ceramic_panel_designs2} together with the results of FEA simulations (red triangles) based on the MLP predictions. One of the best cases for each scenario was selected for FEA simulations (the input parameters for these cases are included in
Tables~\ref{table:first-scenario} and~\ref{table:second-scenario}).
The weighted pairs for the important parameters (edge temperature and out-of-plane deformation for the thermal sink case) were calculated and the results were superimposed with the MLP predictions of the performance of that specific case (e.g. case 1 for the first scenario). The finite element convergence restrictions may affect the choice of the case to evaluate. 
Good agreement between the true FEA results
and the corresponding MLP prediction 
proves the capabilities of deep learning algorithms, in general, and MLP, in particular, in capturing the thermo-mechanical behaviour.
Minor differences in the sum of the weighted pairs between the FEA simulations and MLP may have their origin primarily in the repeatability and mesh dependency of the FEA simulations, and in the capabilities of the MLP model in learning the thermo-mechanical response, Fig.~\ref{fig:method_comp}.
Similarly, the ability of the CNN algorithm in capturing the thermo-mechanical performance
is shown in Fig.~\ref{fig:CNN-best-cases.png}. In this case, the performance of CNN algorithm on the second scenario (thermal sink) using the weighted pair of $25 \%$ and $75\%$ for internal energy and elastic strain energy, respectively, is shown. Similar to the MLP results shown in 
Fig.~\ref{fig:MLP-best-cases.png}, small differences in the sum of weighted pairs of FEA the simulation (red triangles) and the corresponding CNN prediction (case 0 in green circles) 
shows 
the success of the CNN model in predicting the transient response . 

The success of capturing the transient time-dependent response of the system, and finding the optimum design from an enormous pool of potential design targets, 
demonstrates the power 
of deep learning methods.
This opens the door for using these techniques to various other time-dependent engineering problems. 

\section{Conclusions}\label{Conclusion}

Application of deep learning methods based on MLP and CNN to solve time-dependent thermal problems is reported here. The common approach is using stead-state conditions 
and the current study presents a step forward in extending the capabilities of machine learning to capture the transient response of architectured ceramics under thermal shock conditions. 

The thermo-mechanical problem was first modelled by finite element simulations. 
Then, a range of geometric parameters
were selected to study the performance of the different designs under thermal shock conditions. FEA simulation results were used to train the deep learning algorithms using MLP and CNN to predict the transient thermo-mechanical responses of possible panel designs; 
MLP and CNN are two classes of artificial neural networks that have been used for decades in a wide range of applications espcially in image and speech recognition. 

Here, two networks were trained using simulation data from FEA to learn the relation between the geometric parameters and outputs included the safety factor, friction force, internal energy, out-of-plane deformation, edge temperature, heat rate, contact energy, elastic strain energy, and input power over the simulation period.
When the networks were fully trained and tested, they were employed to assist in finding the optimum designs in  two case studies: 1) thermal shielding applications and 2) thermal sink applications. The first scenario involved edge temperature and out-of-plane deformation as the target (output) parameters. In the second case the performance 
measured using internal energy and elastic strain energy. In each of the two above scenarios, the importance of the given target parameters was varied by changing the effective weight factors of that output in the search algorithm. 

The predictions of the MLP and CNN algorithms for each of these scenarios were then evaluated by performing FEA simulations using the best performing predictions. The comparison between performance of ML-proposed ceramic designs with the corresponding FEA simulations demonstrated the capabilities of deep learning algorithms, MLP and CNN in this case, in not only capturing the complex response of transient thermo-mechanical problems, but also significantly reducing the cost in finding the optimum designs from a pool of millions of possible designs.
Furthermore, we showed that the optimized design after neural network prediction improved edge temperature by almost $30\%$. 

\begin{table}
    \centering
    {\includegraphics[width=1\textwidth]{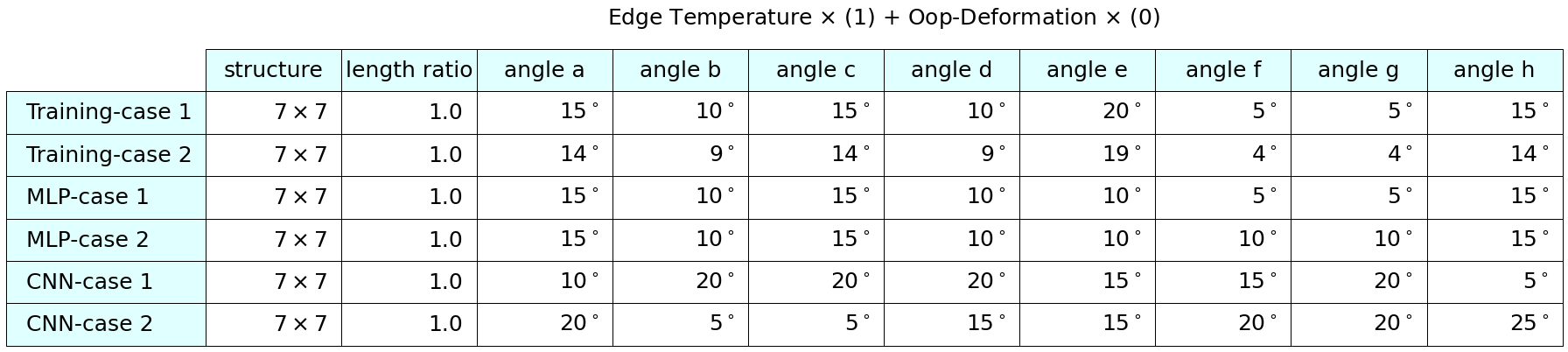}} 
    \caption{First scenario: best performing cases ranked by the lowest maximum edge temperature and lowest maximum out-of-plane deformation during the entire thermal shock cycle.}
    \label{table:first-scenario}
\end{table}

\begin{table}
    \centering
    {\includegraphics[width=1\textwidth]{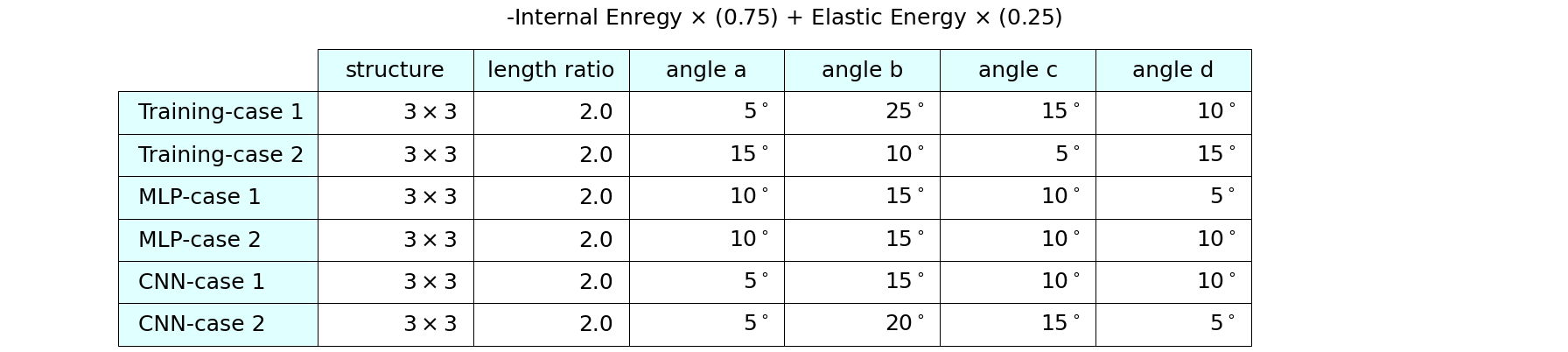}}
    \caption{Second scenario: best performing cases are sorted by a combination of the highest maximum internal energy and lowest maximum elastic strain energy for training, MLP, and CNN predictions.}
    \label{table:second-scenario}
\end{table}

\section*{Acknowledgments}
 The authors thank the new beginning Ideation fund at the National Research Council Canada (NRC). MK thanks the Natural Sciences and Engineering Research Council of Canada (NSERC) and the Canada Research Chairs Program. Computing facilities were provided by SHARCNET (www.sharcnet.ca) and Compute Canada (www.computecanada.ca). HR acknowledges the support from the FRQNT’s B3X postdoctoral fellowship and McGill’s Doctoral Internship Award. 

\section*{Author Contributions}
 E.K. took care of data curation; investigation; methodology; and writing the review and editing.
H.Y. took care of conceptualization; data curation; formal analysis; investigation; methodology; supervision; visualization; roles/writing the original draft; and writing the review and editing. H.R. took care of data curation; investigation; methodology.
R.B. took care of conceptualization; formal analysis; funding acquisition; investigation; methodology; resources; software; supervision; and writing the review and editing. M.R. took care of formal analysis; funding acquisition; investigation; methodology; resources; software; supervision; and writing the review and editing. M.K. took care of formal analysis; funding acquisition; investigation; methodology; resources; software; supervision; and writing the review and editing. B.A. took care of conceptualization; funding acquisition; investigation; methodology; resources; software; supervision; and writing the review and editing.

\section*{Data availability}
The data that supports the findings of this study are
available within the article.

\section*{Declarations}
Conflict of interest The authors have no competing interests to declare
that are relevant to the content of this article.


\end{document}